\documentclass[final]{svjour3}
\usepackage{graphicx}
\graphicspath{{./img/}}
\usepackage{rotating}
\usepackage{amssymb}
\usepackage{mathptmx}
\usepackage{color}
\usepackage[numbers]{natbib}
\bibpunct{[}{]}{,}{n}{}{,}
\makeatletter
\journalname{Journal of Low Temperature Physics}

\newcommand{\micron}{\mu\mathrm{m}}

\usepackage{aas_macros}

\begin{document}

\title{Advanced ACTPol Low Frequency Array: Readout and Characterization of
Prototype 27 and 39 GHz Transition Edge Sensors}

\author{B.J.~Koopman\textsuperscript{1}\kern-1.5pt \and
N.F.~Cothard\textsuperscript{2}\kern-1.5pt \and
S.K.~Choi\textsuperscript{3}\kern-1.5pt \and
K.T.~Crowley\textsuperscript{3}\kern-1.5pt \and
S.M.~Duff\textsuperscript{4}\kern-1.5pt \and
S.W.~Henderson\textsuperscript{5}\kern-1.5pt \and
S.P.~Ho\textsuperscript{3}\kern-1.5pt \and
J.~Hubmayr\textsuperscript{4}\kern-1.5pt \and
P.A.~Gallardo\textsuperscript{1}\kern-1.5pt \and
F.~Nati\textsuperscript{6}\kern-1.5pt \and
M.D.~Niemack\textsuperscript{1}\kern-1.5pt \and
S.M.~Simon\textsuperscript{7}\kern-1.5pt \and
S.T.~Staggs\textsuperscript{3}\kern-1.5pt \and
J.R.~Stevens\textsuperscript{1}\kern-1.5pt \and
E.M.~Vavagiakis\textsuperscript{1}\kern-1.5pt \and
E.J.~Wollack\textsuperscript{8}\kern-1.5pt
}

\institute{\footnotesize
  \noindent\textsuperscript{1}Department of Physics, Cornell University, Ithaca, NY, USA 14853\\
  \noindent\textsuperscript{2}Department of Applied and Engineering Physics, Cornell University, Ithaca, NY, USA 14853\\
  \noindent\textsuperscript{3}Joseph Henry Laboratories of Physics, Jadwin Hall, Princeton University, Princeton, NJ, USA 08544\\
  \noindent\textsuperscript{4}NIST Quantum Devices Group, 325 Broadway Mailcode 817.03, Boulder, CO, USA 80305\\
  \noindent\textsuperscript{5}Kavli Institute for Particle Astrophysics and Cosmology, SLAC National Accelerator Laboratory, 2575 Sand Hill Rd, Menlo Park, California, USA 94025\\
  \noindent\textsuperscript{6}Department of Physics and Astronomy, University of Pennsylvania, 209 South 33rd Street, Philadelphia, PA, USA 19104\\
  \noindent\textsuperscript{7}Department of Physics, University of Michigan, Ann Arbor, USA 48103\\
  \noindent\textsuperscript{8}NASA Goddard Space Flight Center, Greenbelt, MD 20771 USA\\
\email{bjk98@cornell.edu}}

\maketitle

\begin{abstract}

Advanced ACTPol (AdvACT) is a third generation polarization upgrade to the
Atacama Cosmology Telescope, designed to observe the cosmic microwave
background (CMB). AdvACT expands on the 90 and 150 GHz transition edge sensor
(TES) bolometer arrays of the ACT Polarimeter (ACTPol), adding both high
frequency (HF, 150/230 GHz) and low frequency (LF, 27/39 GHz) multichroic
arrays. The addition of the high and low frequency detectors allows for the
characterization of synchrotron and spinning dust emission at the low
frequencies and foreground emission from galactic dust and dusty star forming
galaxies at the high frequencies. The increased spectral coverage of AdvACT
will enable a wide range of CMB science, such as improving constraints on dark
energy, the sum of the neutrino masses, and the existence of primordial
gravitational waves.
The LF array will be the final AdvACT array, replacing one of the MF arrays for
a single season. Prior to the fabrication of the final LF detector array, we
designed and characterized prototype TES bolometers. Detector geometries in
these prototypes are varied in order to inform and optimize the bolometer
designs for the LF array, which requires significantly lower noise levels and
saturation powers (as low as ${\sim}1$ pW) than the higher frequency detectors.
Here we present results from tests of the first LF prototype TES detectors for
AdvACT, including measurements of the saturation power, critical temperature,
thermal conductance and time constants. We also describe the modifications to
the time-division SQUID readout architecture compared to the MF and HF arrays.


\end{abstract}

\section{Introduction}
Advanced ACTPol (AdvACT) is an upgrade to the Atacama Cosmology Telescope
polarimeter (ACTPol), designed to observe the polarization of the cosmic
microwave background (CMB) across five different frequency bands
\cite{2016ApJS..227...21T}. ACT is an off-axis Gregorian telescope located in
the Chilean Atacama Desert at an elevation of 5190 m \cite{Fowler:07}. The
telescope optical chain consists of a 6 m diameter primary mirror, a 2 m
diameter secondary mirror and a set of three optics tubes. Each tube consists
of a window, filter stack, three silicon reimaging optics, a feedhorn array and
finally the detector focal plane. The AdvACT upgrade has already replaced each
of ACTPol's three detector arrays, replacing optics tube elements to
accommodate different frequency bands where needed. The 2017 season
configuration consists of one high frequency (HF) array, observing at 230 and
150 GHz, and two mid frequency (MF) arrays, observing at 150 and 90 GHz
\cite{Henderson2016}. The last array of AdvACT to be deployed is the low
frequency (LF) detector array, which will replace one of the MF arrays and
observe at 27 and 39 GHz. 

Single frequency observations of the CMB are limited by foregrounds such as
synchrotron, spinning dust emission, galactic dust, and dusty star forming
galaxies. AdvACT's high and low frequency coverage allows for the removal of
these foregrounds, with the low frequency coverage aiding in the removal of
synchrotron and spinning dust emission. The wide frequency coverage (27-230
GHz) and fine angular resolution ($1.4'$ at 150 GHz) of AdvACT will enable a
wide range of science such as improving constraints on dark energy, the sum of
the neutrino masses, and the existence of primordial gravitational waves
\cite{Henderson2016}. Here we present the characterization of the detector
test die used to select the final detector parameters for fabricating the
AdvACT LF array.

\section{The Detectors}
The LF array is a polarization-sensitive dichroic array of 73 pixels coupled to
transition edge sensor (TES) bolometers \cite{simon17}. Each pixel has four
orthomode transducer (OMT) coupling probes, two for each linear polarization,
connected to superconducting niobium microstrip lines that transmit radiation
to the AlMn TES bolometers \cite{Li2016, Duff2017LTD}. There are four optical TESes per
pixel, a pair for each linear polarization at both 27 and 39 GHz, as well as
two dark TESes. Each TES island is weakly coupled to the cold bath by a set of
four SiN legs. During operation, the TES is voltage-biased to keep it on the
superconducting transition, which is designed to be at 165 mK. The geometry of
the TES legs determines the thermal conductance to the bath, $G$, which limits
how much power the detector can dissipate before saturating. 

\begin{figure}[htbp]
\begin{center}
\includegraphics[width=0.49\linewidth, keepaspectratio]{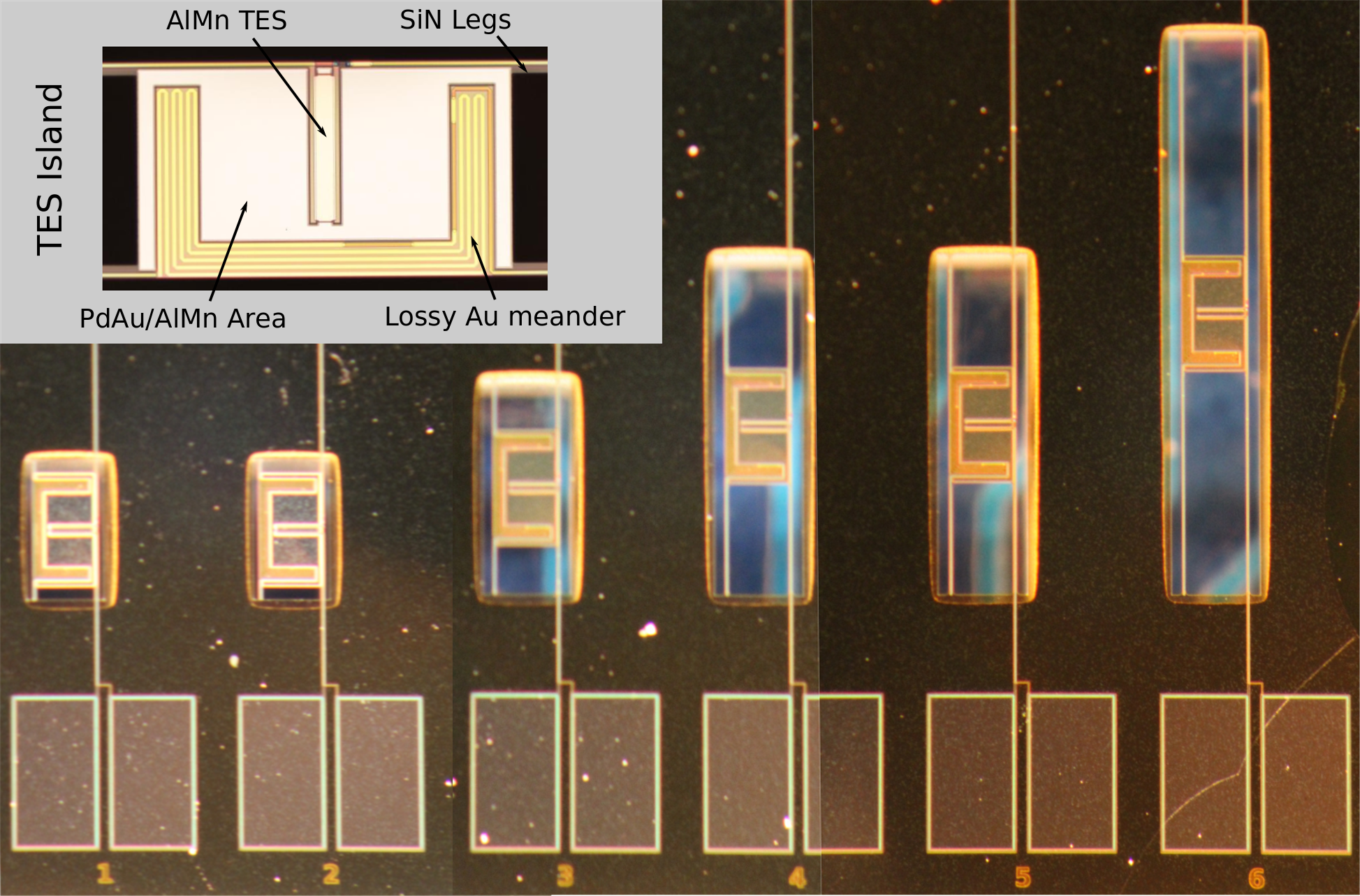}\hfill\includegraphics[width=0.49\linewidth, keepaspectratio, angle=180, origin=c]{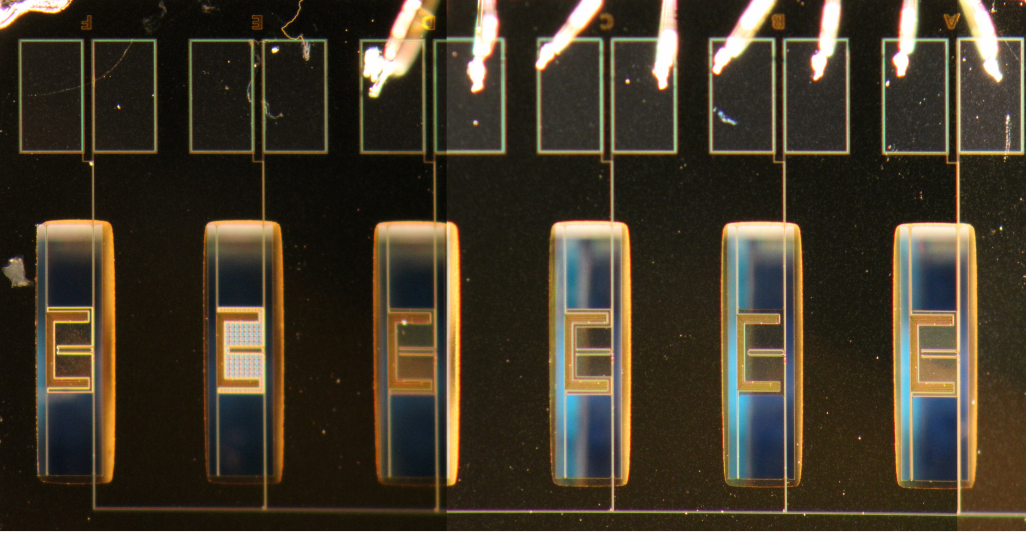}
\caption{(Left): LF prototype transition edge sensors with different leg
geometries. The devices are labeled 1 to 6 from left to right. Table
\ref{table:results} lists the design frequency and leg parameters for each of
the devices. Inset to this figure in the upper left is a close up of a single
labeled TES island. (Right): LF prototype TESes with different PdAu/AlMn
volumes. The devices are labeled A to F from left to right. Devices A through D
have no PdAu, with devices A and D having the full AlMn volume. Device B only
has AlMn in the central region that forms the TES, while device C has a reduced
area of AlMn. Devices F has a full volume of PdAu and AlMn and device E has a
patterned reduction of PdAu, referred to as ``swissed''. (Color figure
online.)}
\label{fig:teses}
\end{center}
\end{figure}

\begin{table}[htbp]
\begin{center}
\begin{tabular}{|c|c|c|c|c|c|c|c|c|}
\hline
ID & $f$ [GHz] & $w$ [$\micron$] & $l$ [$\micron$] & $T_c$[mK] & $P_{sat}$[pW] & $G$[pW/K] & $R_n$[m$\Omega$] & $n$\\
\hline
1 & 39 & 14.4 & 61.0 & $178.9\pm0.1$ & $12.5\pm0.7$ & $276.1\pm15.5$ & $7.9\pm0.1$ & $3.4\pm0.1$\\
2 & 39 & 12.1 & 61.0 & $178.5\pm0.2$ & $10.5\pm0.3$ & $229.3\pm7.5$ & $7.8\pm0.1$ & $3.4\pm0.1$\\
3 & 27 & 10.0 & 219.8 & $173.4\pm1.0$ & $4.1\pm0.2$ & $88.3\pm4.5$ & $7.5\pm0.1$ & $3.0\pm0.1$\\
4 & 27 & 10.0 & 500.0 & $170.5\pm1.2$ & $2.1\pm0.1$ & $44.0\pm1.7$ & $7.5\pm0.1$ & $2.8\pm0.1$\\
5 & 27 & 15.0 & 500.0 & $172.9\pm1.9$ & $3.5\pm0.4$ & $76.5\pm7.3$ & $7.5\pm0.1$ & $3.0\pm0.1$\\
6 & 27 & 10.0 & 1000 & $164.7\pm2.5$ & $1.0\pm0.1$ & $22.6\pm2.3$ & $7.5\pm0.1$ & $2.7\pm0.1$\\
\hline
\end{tabular}
\caption{Measured detector properties. ID corresponds to the device design
variations shown in Fig. \ref{fig:teses} (Left), $f$ is the design frequency
for the detector, $w$ is the TES leg width, $l$ the TES leg
length, $T_c$ is the critical temperature of the TES,
$P_{sat}$ is the saturation power of the detector, $G$ is the
thermal conductivity, $R_n$ is the normal resistance, and $n$ is the
thermal conductivity exponent. Parameters correspond to a bath temperature of
100 mK, which will be the temperature of the devices in the field. The mean and
standard deviation of each parameter for the measured devices is shown. Errors
shown do not account for correlation between parameters. Section
\ref{sec:device_characterization} discusses the correlation between $T_c$ and
$n$.}
\label{table:results}
\end{center}
\end{table}

The TES leg geometry is selected to optimize the performance under expected
loading conditions at the ACT site. This selection is largely driven by the
saturation power without optical loading, $P_{sat}$, and the thermal
conductivity, $G$. The target saturation powers are 1.5 pW at 27 GHz and 7.8 pW
at 39 GHz, which corresponds to three times the estimated loading at each
frequency. These are much lower saturation powers than those in the MF ($12.5$
and $11.3$ pW) and HF ($25$ and $12.5$ pW) arrays, driving us to much longer
leg lengths. However, similarly long legs have been fabricated for SPIDER
\cite{spider}.

The following model is used to describe how $P_{sat}$ is determined by the bath
temperature, $T_{bath}$, and the critical temperature of the device, $T_c$:
\begin{equation}
P_{sat} = K(T_c^n - T_{bath}^n).
\label{psat}
\end{equation}
\noindent The thermal conductivity, $G$, is then given by,
\begin{equation}
G = \frac{dP_{sat}}{dT_c} = nKT_c^{n-1}.
\label{g}
\end{equation}

The LF detector test die leg parameters were selected by extrapolating a linear
fit to the $P_{sat}$ for the MF and HF detectors as a function of the
cross-sectional area to the leg length, $A/l$. Test dies with several leg
variants were then fabricated to explore the parameter space near $P_{sat} =
1.5$ and near $P_{sat} = 7.8 \,\mathrm{pW}$. These variants are shown in
Fig.\ref{fig:teses}. Additional devices with different heat capacities were
fabricated to optimize the temporal response of the detector. The detectors
must respond quickly enough to prevent smearing of the signals from the sky or
from the half-wave-plate modulators \cite{Henderson2016}, but this must be
balanced against the temporal response becoming so fast that the detectors
become unstable \cite{Irwin2005}. Both types of devices were characterized, and
the final leg and heat capacity geometries for use on the LF array were
selected based on those results.

\section{Device Characterization}
\label{sec:device_characterization}

\begin{figure}[bp]
\begin{center}
\includegraphics[width=0.8\linewidth, keepaspectratio]{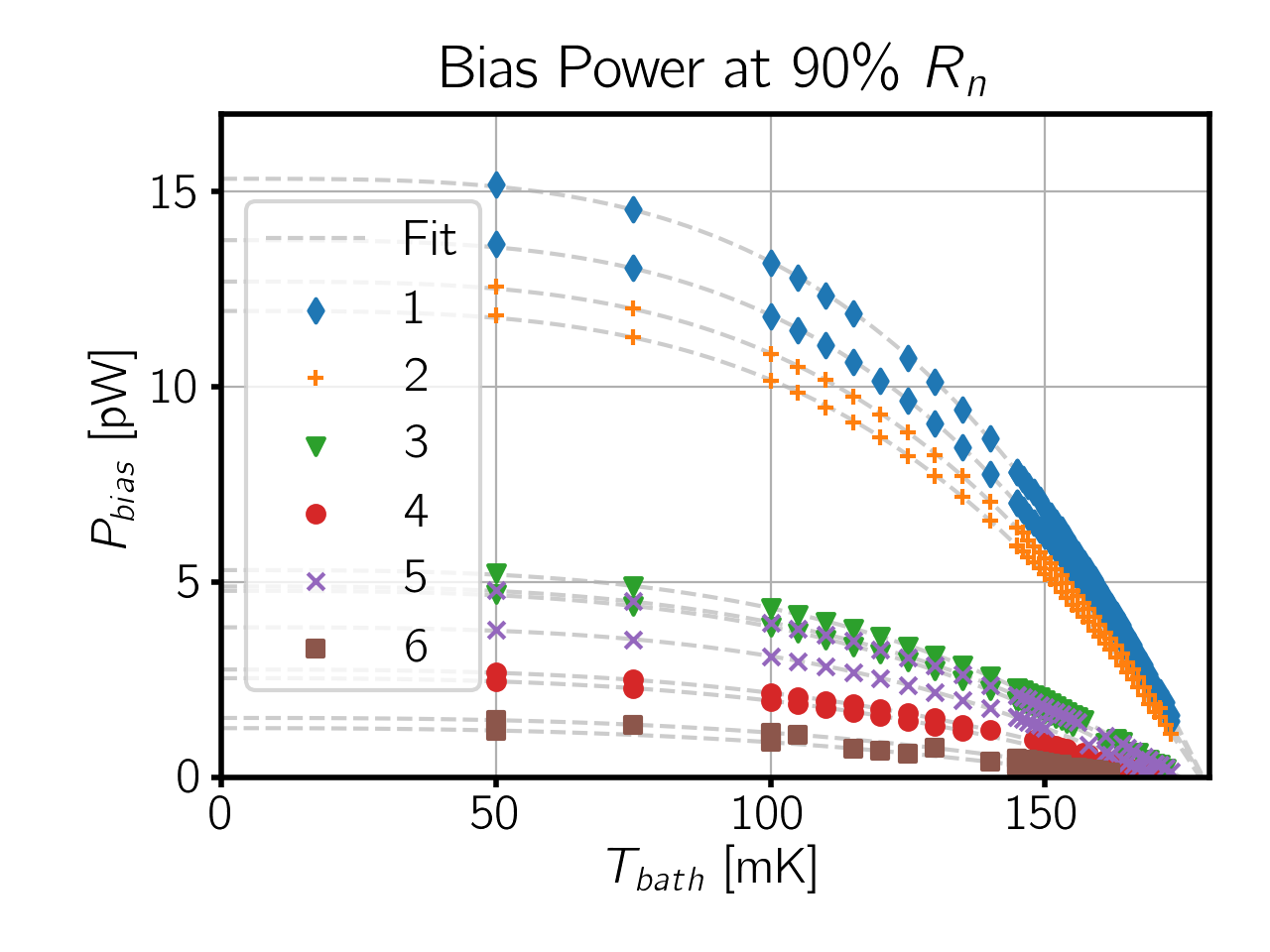}
\caption{Bias power at 90\% the normal resistance, $R_n$, as a
function of the bath temperature, $T_{bath}$, for each of the tested prototype
TES devices. The upper group, devices 1 and 2, are the 39 GHz devices while
the lower group, devices 3, 4, 5, and 6, are lower saturation power 27 GHz
devices. The fit for each individual device is to Eq.(\ref{psat}). (Color
figure online.)}
\label{fig:psat_fits}
\end{center}
\end{figure}

The optimal TES leg geometries were determined by measuring current-voltage
(IV) curves at many different bath temperatures, ranging from 50 to 180 mK.
These IV curves are used to determine the bias power required to drive the TES
to 90\% of the normal resistance, $R_n$. We define this value to be the
saturation power, $P_{sat}$. The bias powers at 90\% $R_n$ for each bath
temperature are fit to Eq. (\ref{psat}), as shown in Fig. \ref{fig:psat_fits},
for $K$, $T_c$, and $n$.

\begin{figure}[htbp]
\begin{center}
\includegraphics[width=0.8\linewidth, keepaspectratio]{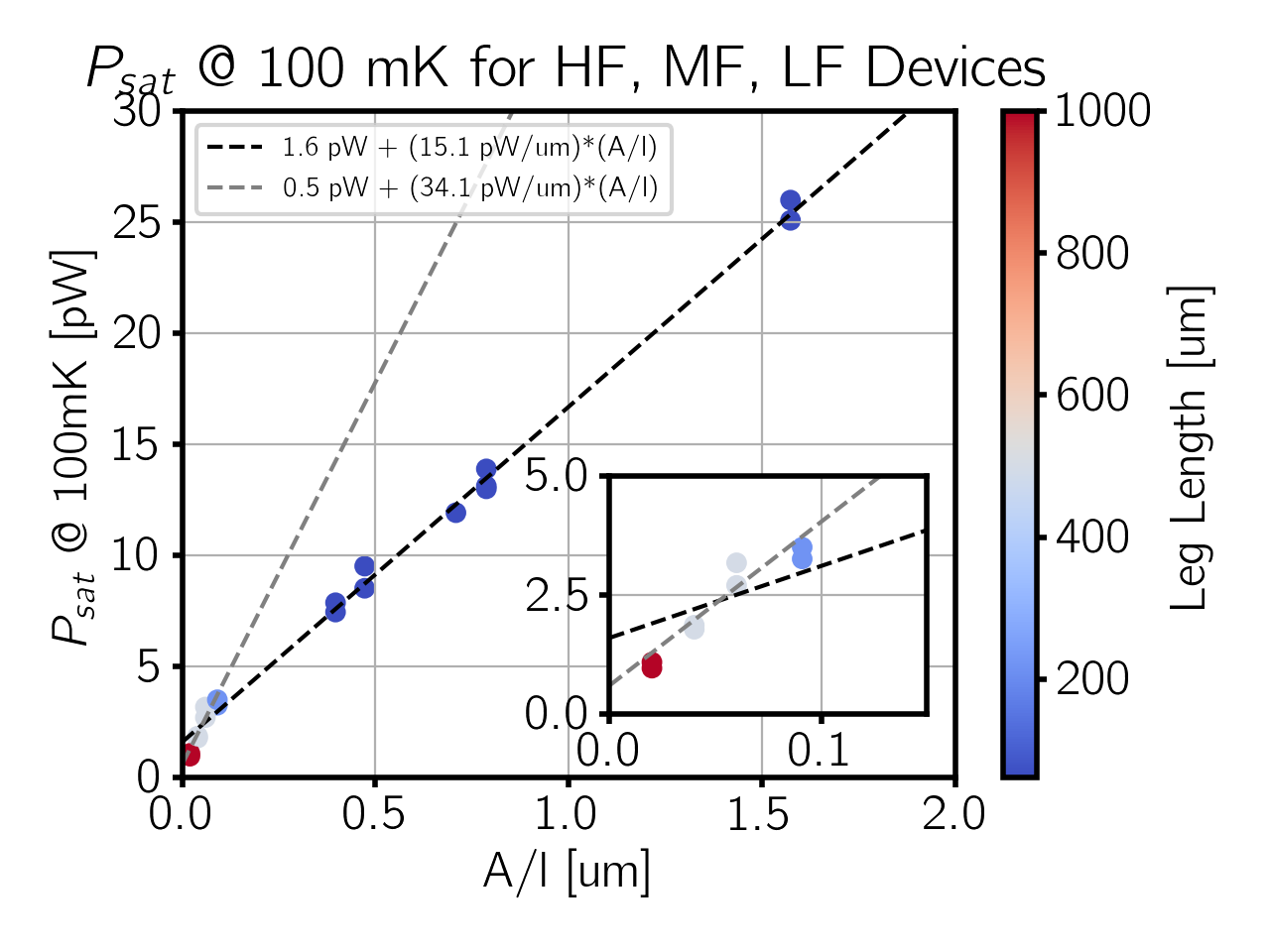}
\caption{Plot of $P_{sat}$ versus $A/l$, where $A$ is the cross-sectional area
of one of the four TES legs and $l$ is its length, for typical HF (230/150
GHz), MF (150/90 GHz), and LF (39/27 GHz) devices. The color scale shows the
leg length, with all HF, MF and 39 GHz LF legs being $61\,\micron$ long, while
the LF 27 GHz test devices have a range of longer leg lengths as shown in Table
\ref{table:results}. The black dashed line fit is to the HF, MF, and 39 GHz LF
detectors. The gray dashed line is a fit to just the 27 GHz LF detectors, which
have the lowest saturation power of all the AdvACT detectors, and differ from
the trend exhibited by the $l = 61.0 \,\micron$ detectors.  We allow for a
non-zero $P_{sat}$ offset in the fit, which may be present due to residual
conductance from the Nb. The inset plot is a zoom in on the low $P_{sat}$ LF
detectors. (Color figure online.)}
\label{fig:hf_mf_lf}
\end{center}
\end{figure}

Based on previous measurements of the $P_{sat}$ values for the MF and HF
detectors, we expect $P_{sat}$ to be proportional to the ratio of $A/l$
\cite{Henderson2016}. In Fig. \ref{fig:hf_mf_lf}, we show the measured
$P_{sat}$ vs the ratio of $A/L$. The black-dashed line is the best-fit linear
relation, which was used to select the leg geometry of the 39 GHz detectors.
For these, we choose to keep the leg length the same as the HF and MF
detectors, $l = 61\,\micron$, which gave a width, $w$, of $12.1\,\micron$.

The 27 GHz detectors have the lowest target $P_{sat}$ of all the AdvACT
detectors at $1.5\,\mathrm{pW}$. While the test die parameters were selected
based on the HF and MF linear fit, we find that the low saturation power,
longer leg length detectors differ in their behavior with $P_{sat}$ as a
function of $A/l$ in that the slope is larger by about a factor of two. This
may be due to the phonon transport differing for the much longer legs. This
difference is reflected in the 27 GHz devices having a thermal conductance
exponent, $n$, $\leq3$, compared to $n \sim 3.4$ for the 39 GHz devices, as
shown in Table \ref{table:results}. While we do not account for correlation
between fit parameters in our error determination in Table \ref{table:results},
we do find there to be a negative correlation between $T_c$ and $n$, so for a
fixed $T_c$ the difference between the 27 and 39 GHz detector $n$ values will
persist. We fit the devices with $l > 61\,\micron$ independently, leading to
the gray dashed line in Fig.\ref{fig:hf_mf_lf}, which we used to select the
final 27 GHz detector leg parameters of $w = 10\,\micron$ and $l =
628\,\micron$. Measured prototype device parameters are shown in Table
\ref{table:results}. The measured critical temperatures for most of the devices
were found to be above the target 165 mK.  Selection of the final leg
geometries for fabricating the LF array was done after scaling these results to
a $T_c$ of 165 mK.

The detector responsivity decreases with increasing frequency due to the
thermal time constant of the detector, which can be varied by adding heat
capacity to the TES island. We tested six different recipes of PdAu and
AlMn, in order to explore how the time constant varied as a function of each
materials volume. These six test devices were all 27 GHz devices, four with
AlMn volumes ranging from $2639.6\,\micron^3$ to $36538.4\,\micron^3$ and no
PdAu, two of which had the same AlMn volume but different leg widths, and two
with the full $36538.4\,\micron^3$ AlMn volume while having either
$21451.8\,\micron^3$ or $36112.8\,\micron^3$ of PdAu. Extrapolations from the
previously fabricated MF devices were suitable for the 39 GHz devices.

We report the time constant of the detector as $f_{3dB} = 1/2
\pi \tau$, defined as the frequency at which the response of the detector
decreases by a factor of two compared to the DC response. The target $f_{3dB}$
for the LF detectors ranges from 81 to 275 Hz for 35\%-70\% $R_n$ across the
range of expected bias powers under loading in the field. This range is set by
requiring fast enough detectors to observe with a rapidly rotating half-wave
plate modulator, but not too fast that they could become unstable \cite{Henderson2016}.

\begin{figure}[htbp]
\begin{center}
\includegraphics[width=0.8\linewidth, keepaspectratio]{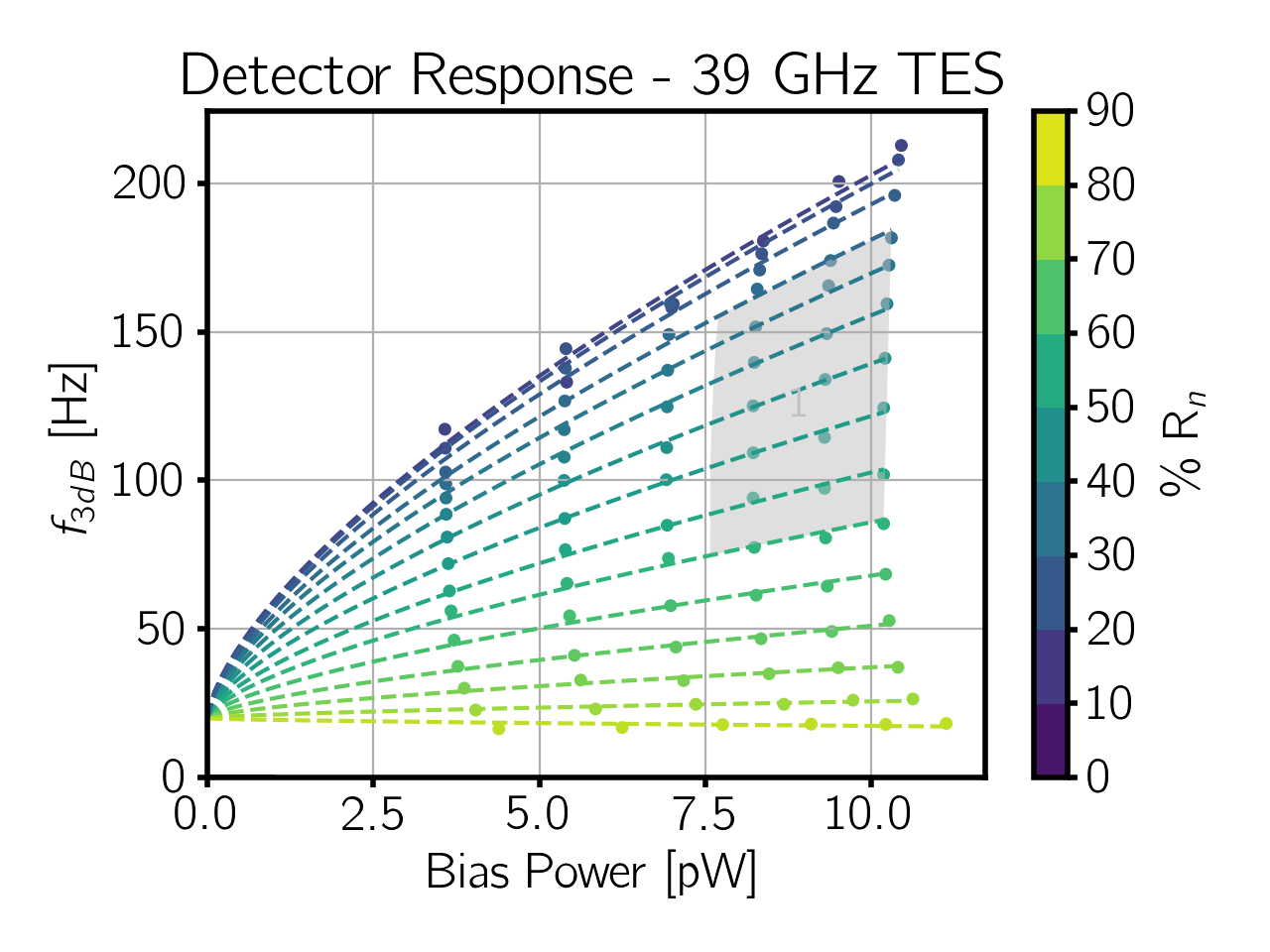}
\caption{Bias step results for a single 39 GHz prototype detector. The gray
region indicates the targeted operating range of the detector in the field.
Each vertical grouping of points is at one bath temperature. The bias point in
\%$R_n$ is selected and then a square wave is applied. The $f_{3dB}$ is
measured from the response to the square wave and this is repeated for a full
range of bias points before moving to the next temperature. These data are used
to estimate the detector response speed under a range of optical loading
conditions in the field. (Color figure online.)}
\label{fig:250khz}
\end{center}
\end{figure}

We test the response time by adding a small square-wave to the voltage bias
applied to the detectors with amplitude ranging from $1-5\%$ of the DC bias
value. The response is sampled at 250 kHz to measure the exponential fall time
of the detector current, or $\tau$. We do this at six different bath
temperatures and several different points on the TES transition, measured by
the TES resistance relative to normal, $\%R_n$, as shown in Fig.
\ref{fig:250khz}. These data are then fit to a two-fluid model described by Eq.
(\ref{twofluid1}), where $A$ and $B$ are a function of measurable parameters
specified in \cite{doi:10.1063/1.367153}, and $P_{bias}$ is the bias power,
\begin{equation}
    f_{3dB} = A + B P_{bias}^{2/3}.
\label{twofluid1}
\end{equation}
This model qualitatively captures the data, though a linear
model yields similar results. Better understanding of the TES behavior and
physics throughout the transition is work in progress.

From these data, we found that having PdAu on the 27 GHz devices, in addition to
the full AlMn volume, decreased the $f_{3dB}$ below our target threshold of 81
Hz. Based on this we chose to have 73\% of the full (largest tested) AlMn
volume (full being $36538.4\,\micron^3$) and no PdAu on the 27 GHz detectors
and 100\% of the full AlMn volume and 76\% of the full volume of PdAu (full
being $36112.8\,\micron^3$) for the 39 GHz detectors.

\section{Readout}
The LF array will have a much smaller number of TESes than AdvACT's HF and MF
arrays but it will be read out using the same time division multiplexing (TDM)
scheme \cite{Henderson2016, 2016SPIE.9914E..1GH, Crowley2017LTD, Choi2017LTD}.  Each TES in the LF array
will be voltage-biased and multiplexed through the warm Multi-Channel
Electronics (MCE) using a SQUID-based TDM architecture developed at
NIST/Boulder \cite{2008JLTP..151..908B, 2008SuScT..21j5022B}.  Containing fewer
pixels, the LF cold electronics will be a simplified adaptation of the HF and
MF designs, using the same PCB used in the HF and MF arrays. The LF array will
have a multiplexing (MUX) factor, or number of detectors per readout channel,
of 26:1, whereas HF and MF have MUX factors of 64:1 and 55:1, respectively. The
cold readout electronics are similar otherwise to the electronics described in
\cite{Henderson2016, 2016SPIE.9914E..1GH}.

The LF array readout wiring has been designed such that TESes from polarization
pairs at one optical frequency are read out on the same column and thus the
same 1 K SQUID series array and 300 K warm amplifier. Unlike in the MF and HF
arrays, each column in the LF array has its own dedicated TES bias line and
dark TESes are wired to their own column. This is an improvement over the MF and
HF readout because each detector type (27 GHz, 39 GHz, and dark), which will have
different optimal bias powers, can be independently biased. Deliberate shorts
on some unused shunt inputs have also been implemented, providing Johnson noise
channels for independently probing detector bias line noise.

\section{Conclusion}

We fully characterized new test dies, which were designed to demonstrate the
low $P_{sat}$ values needed to optimize the LF array. The LF cryogenic readout
electronics have been assembled and tested. The LF detector array is currently
being fabricated at NIST and will be deployed to extend the range of AdvACT
from 27 GHz to 230 GHz.

This demonstrates our ability to cover a wide range of saturation powers (as
low as 1 pW) and detector response speeds through modification of the detector
leg length, the leg width, the AlMn volume, and the PdAu volume on a single
detector design. This has direct application to detector design for future CMB
experiments, such as the Simons Observatory and CMB-S4, which will face similar
design challenges.

\begin{acknowledgements}
This work was supported by the U.S. National Science Foundation through award
1440226. The development of multichroic detectors and lenses was supported by
NASA grants NNX13AE56G and NNX14AB58G. The work of KPC, KTC, BJK, and JTW was
supported by NASA Space Technology Research Fellowship awards. 
\end{acknowledgements}

\bibliography{ltd}
\bibliographystyle{ltd16}

\end{document}